\documentclass[iop]{emulateapj}
\usepackage{enumerate,color}

\shorttitle{The Scaling of Inner  Mass with Stellar Mass}
\shortauthors{Kuzio de Naray \& McGaugh}

\begin{document}
\submitted{Accepted to ApJ Letters}
\title{The Inner Dynamical Mass Across Galaxy Morphology:\\ A Weak Scaling with Total Stellar Mass}
\author{Rachel Kuzio de Naray}
\affil{Department of Physics \& Astronomy, Georgia State University, P.O. Box 5060, Atlanta, GA 30302-5060, USA}
\author{Stacy S.~McGaugh}
\affil{Department of Astronomy, Case Western Reserve University, Cleveland, OH 44106, USA}
\email{kuzio@astro.gsu.edu}
\email{stacy.mcgaugh@case.edu}

\begin{abstract}
We compile a sample of dispersion-supported and rotation-supported galaxies for which the dynamical mass enclosed at 500 pc can be measured.  We find that this dynamical quantity increases only slowly with stellar mass ($M_{total}(<500\mathrm{pc}) \propto M_{stars}^{0.16}$) over $\sim$~9 decades in baryonic mass and $\sim$~3 in length scale, with a sudden upturn at the highest masses ($M_{stars} \gtrsim 10^{10}\;\mathrm{M}_{\sun}$). This upturn occurs for the earliest type (S0 and Sa) disk galaxies, and is consistent with the ``extra'' mass within 500 pc being predominantly stellar.  This sudden change may be indicative of different bulge formation mechanisms between early and late type disks.  We also compare the data to the results of recent galaxy formation simulations.  These broadly track the observed trend, with some tendency to put too much mass in the inner 500 pc, depending on the details of each simulation.
\end{abstract}


\section{Introduction}
\label{intro}
Galaxy formation remains one of the more vexing problems in extragalactic astronomy.
As shown in recent papers such as \citet{Brook2012}, \citet{Governato2012}, \citet{DiCintio2013}, and \citet{Vogelsberger2013}, galaxy simulations that aim to reproduce the observed properties of galaxies and their dark matter halos over a wide range of masses have become increasingly complex and comprehensive.  Simulations are probing, among other topics, the seeding and growth of black holes, the chemical enrichment of galaxies, the formation of bulgeless disk galaxies, and the ability of stellar feedback to modify the inner structure of dark matter halos.  The simulations often tackle more than one of these issues simultaneously.

To evaluate how realistic the modeled formation processes and resulting simulated galaxies are, having a well-studied and understood sample of real, observed galaxies with which to compare is necessary.  Galaxy observations provide important constraints on the initial input parameters to the simulations and also define the final galactic structure and composition that should be achieved.

In this Letter, we compile data for a sample of dwarf spheroidal and early and late-type disk galaxies and characterize their inner mass structure.  We investigate the amount of total enclosed dynamical mass inside of 500~pc across galaxy type and total baryonic mass.  We also directly compare recent galaxy simulations to the data to determine how well they reproduce observed galaxy structure.

\section{Data} 
\label{data}
We collect data for a sample of 50 dispersion- and 61 rotation-supported systems.  The dispersion-supported systems are Local Group dwarf spheroidal (dSph) galaxies with measured half-light radii and stellar velocity dispersions.  The data are taken from \citet{Walker2007}, \citet{McConnachie2012}, and \citet{Collins2013}.  These systems have total luminosities ranging from $\sim$~10$^{3}$$-$10$^{7}$ $L_{\odot}$.  

The rotation-supported systems are disk galaxies of all morphological types with published rotation curves from \citet{Begum2004}, \citet{deBlok2002}, \citet{deBlok2001}, \citet{deBlok2008}, \citet{Kuzio2008}, \citet{Noordermeer2007}, \citet{Oh2011}, \citet{Sanders1996}, \citet{Sofue1999}, \citet{Trachternach2009}, and \citet{Verheijen2001}.  These include bright, early-type spirals and low mass dwarf irregulars, and both high and low surface brightness galaxies.  The sample of disk galaxies covers a large range in mass, with $V_{flat}$ ranging from $\sim$~20$-$300~km s$^{-1}$.  The unifying property of the sample is the availability of dynamical data at small radii.

For each dSph and disk galaxy, we compile $M_{\rm stars}$, $M_{\rm gas}$, $M_{\rm baryons}$, $R_{\rm e}$, and $M_{\rm total}$($<$500~pc).  For the disk galaxy sample, we also compile the disk type using the HyperLeda database \citep{Paturel2003}.  $M_{\rm stars}$ is the total stellar mass of a galaxy.  We assume a stellar mass-to-light ratio $\Upsilon_{*}$~=~1 when computing $M_{\rm stars}$ for the dSphs \citep[][]{Martin2008}.  For the disk galaxies, we calculate the total stellar mass using the Baryonic Tully-Fisher Relation (BTFR) of \citet{McGaugh2012}.  For $M_{\rm gas}$, 1.33$M_{\rm HI}$ is taken as the total gas mass of a galaxy.   $M_{\rm baryons}$, the total baryonic mass of a galaxy, is the sum of $M_{\rm stars}$ and $M_{\rm gas}$.  The effective radius, $R_{\rm e}$, is the half-light radius in $V$ for the dSph sample, and 1.7 times the disk scale length in $B$ or $R$ for the disk galaxy sample.  

We measure the total dynamical mass (stars+gas+dark matter) $M_{\rm total}$($<$500~pc) inside of 500~pc.  Our primary motivation for choosing 500~pc is so that we can evaluate dSph and disk galaxies at the same scale.  As discussed in detail in Section~\ref{results}, this radius is a compromise for both types of galaxies:  the half-light radius for  dSphs is often less than 500~pc, but resolution and non-circular motions can affect rotation curves at smaller radii.  

To calculate $M_{\rm total}$($<$500~pc) for the dispersion-supported dSph galaxies, we use Equation 10 of \citet{Walker2009}
\begin{equation}
M(r) = \frac{5 r_{\rm half} \sigma^{2} (\frac{r}{r_{\rm half}})^{3}}{G [1 + (\frac{r}{r_{\rm half}})^{2}]},
\end{equation}
and evaluate it at r~=~500~pc.

As discussed in the \citet{Walker2009} paper, this equation provides enclosed mass estimates for dispersion-supported galaxies that are comparable to the masses obtained from a full Jeans/MCMC analysis if it is applied ``near" the half-light radius. The half-light radii of the dSphs in our sample range from tens of pc to $\sim$~1~kpc, with the majority being around 250~pc.  We discuss the applicability/implementation of Equation~1 to our dSph sample below.

For eight  ``classical" dSphs in our sample, $M_{\rm total}$($<$500~pc) is provided by M.G.~Walker (2013, private communication).  These values are calculated using a complete Jeans/MCMC analysis.  We use these data to determine a correction factor, $f$, that we apply to the $M_{\rm total}$($<$500~pc) values calculated for the dSph sample using Equation~1.  This correction factor is the ratio of the fully modeled mass to the mass estimated by Equation~1 as a function of the difference between 500~pc and the half-light radius: \begin{math}f = 0.12\ (0.500 - r_{\rm half}) + 0.78.\end{math}  We discuss in Section~\ref{M500} the typical size of this correction and the uncertainty of $M_{\rm total}$($<$500~pc) values calculated for the dSphs\footnote{We attempted a similar exercise for giant ellipticals, but lack sufficient information to construct the equivalent of Equation 1.} using Equation~1.
 
We lack detailed mass models for some of the assembled disk galaxies, so we assume spherical symmetry: $M(r)=V_{rot}^{2}(r)\;r/G$ with $r=500$~pc. This is a tolerable assumption for very late types, which tend to be dominated by their dark matter halos, and for very early types in which the bulge component dominates the inner mass budget.  For intermediate types, the stellar disk can dominate the mass budget at small radii, but deviations from the spherical approximation are at most tens of percent in mass, a negligible amount over the scales considered.

\begin{figure*}[ht]
\epsscale{0.8}
\plotone{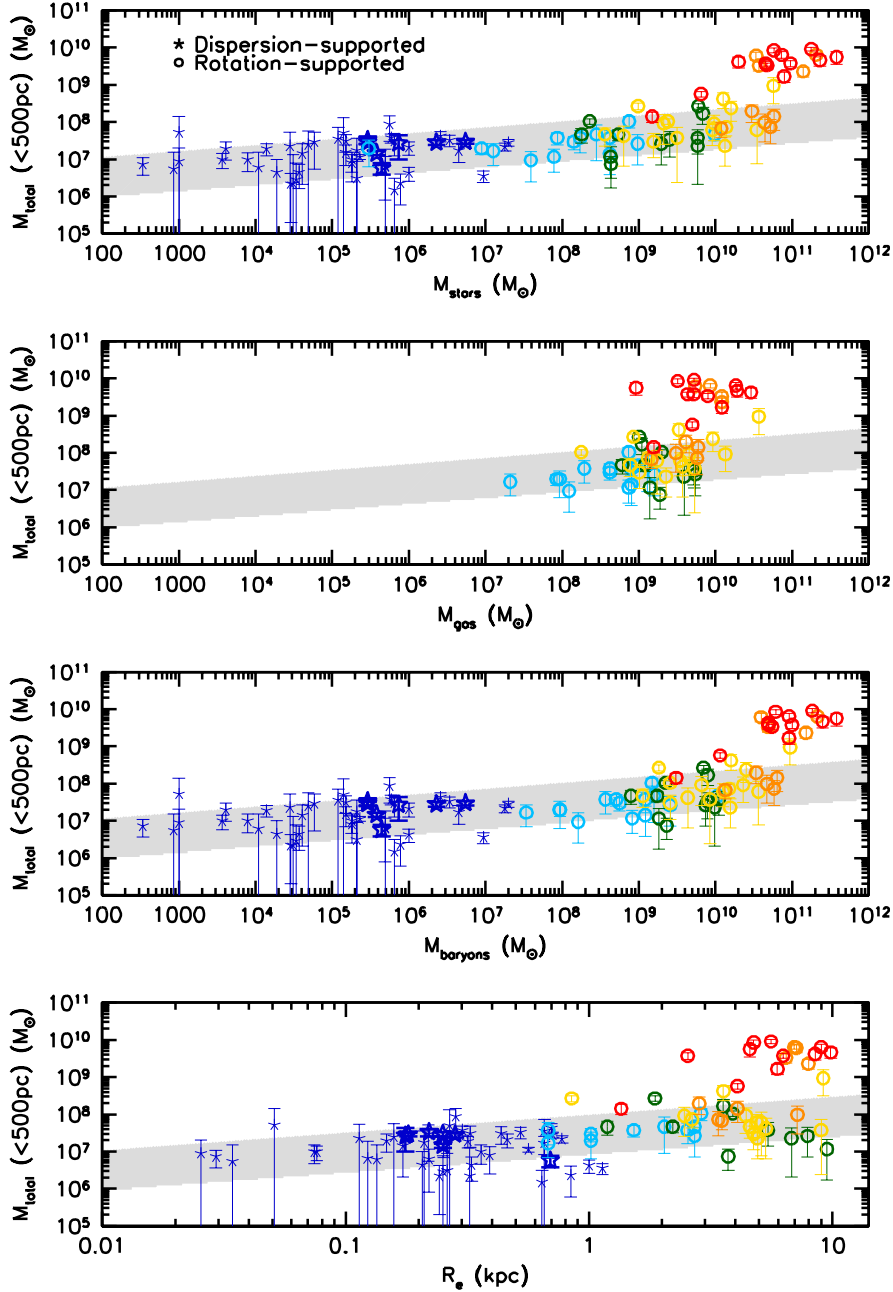}
\caption{Total dynamical mass (stars+gas+dark matter) inside of 500~pc plotted against, from top to bottom, total stellar mass, total gas mass, total baryonic mass, and effective radius.  Dispersion-supported dSph systems are plotted as dark blue stars, with the open stars representing the data from Walker (2013, private communication).  Rotation-supported disk galaxies are plotted as open circles, with the following color/galaxy-type combinations:  red/S0-Sa, orange/Sb-Sbc, yellow/Sc-Scd, green/Sd-Sdm, light blue/Sm-Irr.   Over $\sim$~9 orders of magnitude in baryonic mass and 3 in length, the mass inside of 500~pc increases less than 1.5 orders of magnitude (the gray band).  The relation in the top panel between total inner dynamical mass and total stellar mass is log~$M_{500}$~=~0.16~log~$M_{stars}$~+~6.19 with the width of the gray band set to 1 dex for illustration.  (A color version of this figure is available in the online journal.)}
\label{plot1}
\end{figure*}

\section{Results}
\label{results}

\subsection{Total Enclosed Mass at 500~pc}
\label{M500}
\begin{figure}
\epsscale{1.25}
\plotone{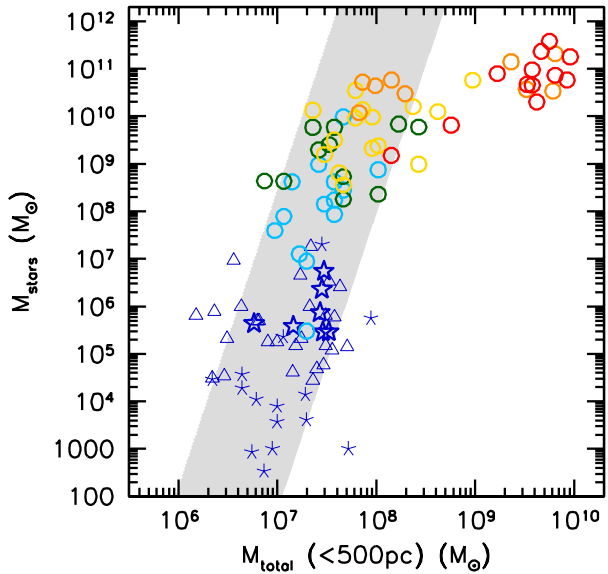}
\caption{Same as the top panel of Figure~\ref{plot1} except that the axes have been flipped, highlighting the spread in $M_{total}$($<$~500~pc).  The Andromeda dSphs are now shown as triangles.  For clarity, error bars are not shown.  Note the apparent gap for the most massive, early type rotators.  It is unclear whether this gap is real, or a chance consequence of the sample size. The sudden increase in central mass is associated with dominant stellar bulges.  (A color version of this figure is available in the online journal.)}
\label{plot2}
\end{figure}

In Figure~\ref{plot1}, we plot $M_{\rm total}$($<$500~pc) as a function of $M_{\rm stars}$, $M_{\rm gas}$, $M_{\rm baryons}$, and $R_{\rm e}$ for the combined dSph and disk galaxy samples.  We find that $M_{\rm total}$($<$500~pc) increases very slowly over many orders of magnitude in the other parameters, and that there is a smooth transition between dispersion-supported systems and rotation-supported systems.  Specifically, over $\sim 9$ orders of magnitude in baryonic mass and $\sim 3$ in length, the total dynamical mass inside of 500~pc increases only 1.5 orders of magnitude for all but the most massive early-type (S0-Sa) disk galaxies. $M_{\rm total}$($<$500~pc) increases from a few~$\times$10$^{6}$~M$_{\odot}$ for the lowest luminosity dSph galaxies to $\sim 10^{8}\;M_{\odot}$ for Sb-Sc type disk galaxies in the sample (see also Figure~\ref{plot2}).  

The dynamical mass enclosed by 500 pc is remarkably similar for vastly different galaxies. Figure~\ref{plot1} shows that galaxies with different kinds of internal motions (pressure support vs.\ rotational support), different primary baryonic mass components (stellar mass  vs.\ gas mass), different environments (satellite galaxies vs.\ isolated field galaxies), different physical sizes, etc., are generally rather similar.  They are not identical as we do find a gradual trend for the inner dynamical mass to increase with total stellar mass: $\log M_{total}(<500\mathrm{pc}) = 0.16 \log M_{stars} +6.19$.  There is some scatter about this relation, but the majority of galaxies fall within half a dex of this line.

The weak correlation defined by the majority of the assembled sample is violated by the most massive early-type (S0-Sa) disks.  These have a significantly higher central concentration of mass (Fig.~\ref{plot2}).  This sets in around $M_{stars} \gtrsim 10^{10}\;M_{\sun}$ with a fairly narrow transition region where galaxies of this mass either lie on or above the relation.  Indeed, the data give the visual impression of a gap between the most concentrated early-type galaxies and the rest of the sample.  It is unclear whether this gap is real, or a chance consequence of the sample size.  The sudden change in enclosed central mass of the earliest type galaxies in our sample is real.

The large central masses of early-type galaxies are probably dominated by stars.  This is consistent with the findings of \citet{Noordermeer2007}, from whom most of the early-type data originate.  In detailed mass models of early-type systems, it is common to see an inner peak in the rotation curve due to the bulge and a subsequent peak farther out.  The latter might be attributed to either disk or dark matter, but one dark matter halo cannot produce both peaks, requiring the inner peak to be dominated by the stellar bulge.

The apparent dichotomy in inner mass is suggestive of distinct bulge formation mechanisms between early- and late-type spirals.  The mechanism in early-type spirals is evidently very efficient at packing lots of stars into a small radius. Perhaps these are classical $r^{1/4}$ bulges built by mergers, or cold flows are more efficient at delivering baryons to small radii in these massive galaxies \citep{Dekel}.  Contrawise, the less dense central regions of late-type galaxies might grow by secular evolution of the disk.

Our results are robust against errors and changes in our assumptions. Uncertainties in $\Upsilon_{*}$ for the dSphs or the calibration of the BTFR for the disks will change $M_{\rm stars}$ by a factor of a few at most.  This translates to small shifts left or right in Figure~\ref{plot1} that are insignificant compared to the entire range of $M_{\rm stars}$ that is plotted, particularly given the shallow slope of $M_{\rm total}$($<$500~pc)  with total stellar mass.  Uncertainties in $M_{\rm gas}$ and $R_{\rm e}$ will produce similarly negligible horizontal shifts in Figure~\ref{plot1}. 

The largest uncertainty is in the measurement of $M_{\rm total}$($<$500~pc).  As discussed in Section~\ref{data}, Equation~1 is a good substitute for full modeling when it is used to determine the enclosed mass of a dispersion-supported system at a radius close to the half-light radius.    Comparing the fully-modeled masses of dSphs provided by Walker to the estimates obtained using Equation~1, we find that Equation~1 gives values that are typically 1.3 times larger than those determined from the full modeling.  This would be a small vertical shift in Figure~\ref{plot1}. Our correction factor, $f$, compensates for this difference when $M_{\rm total}$($<$500~pc) is calculated for the remainder of the dSph sample.  Those dSphs with half-light radii beyond the range of the Walker sample ($\sim$~180$-$700~pc) are least certain. These are typically the lowest luminosity (lowest $M_{\rm stars}$) systems.  Even if we exclude these galaxies, the remaining dSph sample continues the trend set by the disk galaxies, as highlighted by the gray band in the figures.  

The uncertainty on $M_{\rm total}$($<$500~pc) for the disk galaxies is dependent on how well the rotation velocity at 500~pc can be measured.  500~pc is at the lower limit of the spatial resolution of most rotation curves and non-circular motions become more important at small radii.  In addition, it becomes more difficult to constrain $V_{\rm rot}$(500~pc) if the galaxy has a bulge since the rotation curve will rise very quickly.  We are therefore forced to exclude galaxies whose rotation curves lack the rising portion at small radius.  These are typically very massive galaxies with $V_{flat}$ in excess of 300 km s$^{-1}$.  These systems would populate the very upper right corner of Figure~\ref{plot1}. 

Much like the constant mass within 300~pc in the Milky Way dSphs that was emphasized by \citet{Strigari2008}, the total dynamical mass inside of 500~pc appears to be nearly constant in galaxies spanning an enormous range of total baryonic mass, type, and size.  It is only in the earliest type disk galaxies with total stellar masses greater than $\sim$~10$^{10}$M$_{\odot}$ that a significant amount of mass builds up in the inner 500~pc.  In the next section, we explore whether recent galaxy simulations are successful in matching the inner regions of observed galaxies.

\begin{figure*}
\epsscale{0.8}
\plotone{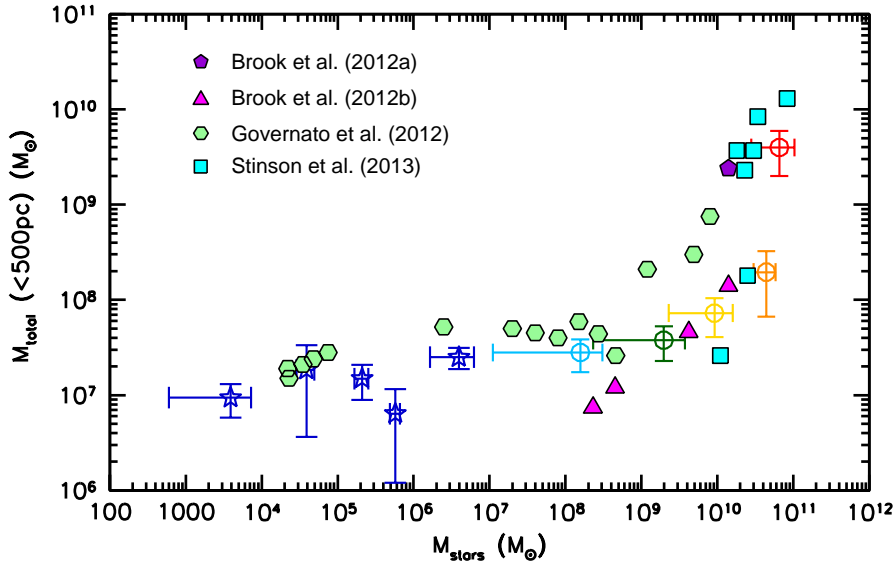}
\caption{Comparison of the total dynamical mass inside of 500~pc in recent galaxy simulations to observed dSph and disk galaxies.  The dSph data in Figures~\ref{plot1} and \ref{plot2} have been binned by total stellar mass with each bin containing 10 objects; the disks have been binned in stellar mass by type with the same color scheme as Figures~\ref{plot1} and \ref{plot2}. The median stellar mass and median inner dynamical mass are plotted for each galaxy bin with error bars showing the median average deviation of each quantity.  The purple pentagon and magenta triangles are the simulations presented in \citet{Brook2012,Brook2012b}, the light green hexagons are the simulations presented in \citet{Governato2012}, and the cyan squares are the simulations of \citet{Stinson2013}.      (A color version of this figure is available in the online journal.) }
\label{plot3}
\end{figure*}

\subsection{Enclosed Mass and Feedback in Simulations}
\label{Fabio}
In Figure~\ref{plot3}, we bin the observed galaxy data and plot the median stellar mass and median $M_{\rm total}$($<$500~pc) for each bin.  The dSph galaxies are binned by stellar mass such that each bin contains 10 galaxies.  The disk galaxies are binned in stellar mass  by type and contain between 9 and 15 galaxies per bin.  Around a few times 10$^{9}$~M$_{\odot}$ in total stellar mass, a gradual upturn in the total dynamical mass inside of 500~pc begins to appear.  This upturn is roughly where the baryons begin to dominate the inner mass \citep{McGaugh2007,Walker2010}. 

We also plot in Figure~\ref{plot3} the total dynamical mass inside of 500~pc for recent galaxy simulations.  These simulations employ feedback prescriptions to modify the inner mass distribution of the galaxy-halo system, prevent the formation of massive bulges, and produce flat rotation curves, exponential surface brightness profiles and dark matter cores.  Specifically, we include the simulations of \citet{Brook2012, Brook2012b}, the simulated galaxies presented in \citet{Governato2012}, and the simulations of \citet{Stinson2013}.    We have chosen these simulations as a comparison sample because $V_{\rm rot}$(500~pc) is measurable in the published rotation curves (or $M_{\rm total}$($<$500~pc) is tabulated) and $M_{\rm stars}$ is also provided.  Not all recent simulations provide the necessary information to perform our test \citep[e.g.,][]{Guedes2011,Hopkins2012,Agertz2013}.

Though $M_{\rm total}$($<$500~pc) does not distinguish between luminous and dark mass, the binned galaxy observations do provide a robust measure of the \textit{total} inner mass.  $M_{\rm total}$($<$500~pc) can therefore be used to gauge how realistic and/or successful the simulations are in reproducing observed total mass distributions.  If they are successful in this regard, one can then ask if they are also able to match the ratios and radial distributions of the different mass components.  With this in mind, we now examine how the simulations compare to the observations in Figure~\ref{plot3}.

The total mass inside of  500~pc in the simulations with $M_{\rm stars}$~$<$10$^{9}$~M$_{\odot}$ are in the same range of the central masses of  observed galaxies of similar total stellar mass, though typically on the high side.  The feedback processes that are acting in these simulations are keeping the total inner dynamical masses consistent with observations.  These simulations also seem to be making progress in matching the mass profile (cuspy vs.~cored) of the dark matter halo (see figure 1 in Governato et al. 2012 and figure 6 in Brook et al. 2012b).

The simulations with $M_{\rm stars}$~$>$~10$^{9}$~M$_{\odot}$ show a similar upturn in $M_{\rm total}$($<$500~pc) as the observed galaxies.  While the \citet{Governato2012} simulations in this range have significantly \textit{more} total mass in their central 500~pc than the observed late-type Sd and Sc disk galaxies with comparable total stellar masses, the two most massive \citet{Brook2012b} simulations are in much better agreement with the observations.  These are their SG3 and SG4 simulations; most of the merger activity and star formation takes place prior to $z$~=~2 in SG3 (the less massive galaxy of the two), while SG4 builds up its mass at a much later time after $z$~=~1.  

The \citet{Brook2012} simulation, identified as an SBc/d galaxy, has significantly more mass in its center than observed galaxies of similar type.   This is not an uncommon result.  Simulated disk galaxies often have very steep rotation curves, even when star formation, feedback and outflows are modeled \citep[see, for example,][]{Guedes2011,Piontek2011}.  The decomposition of the \citet{Brook2012} rotation curve into stellar and dark matter mass components shows that it is not just a build-up of baryonic mass, but also a substantial increase in dark matter mass that is responsible for the large total mass at the center:  the simulated galaxy contains $\sim$~3~$\times$~10$^{8}$~M$_{\odot}$ of dark matter and 2.1~$\times$~10$^{9}$~M$_{\odot}$ in stars in the inner 500~pc.  It is likely that adiabatic contraction has played a role in creating such a dense center.  

The two  \citet{Stinson2013} simulations with low inner dynamical masses fall in the range of observed Sb and Sc galaxies.  These simulations utilize high early stellar feedback and high diffusion to regulate/modify star formation.  The five other \citet{Stinson2013} simulations have much higher inner dynamical masses, more similar to the bin of early-type S0-Sa galaxies.  In general, these simulated galaxies have slightly lower total stellar masses than the observed early-type galaxies.  Alternatively, if the morphology of the simulated galaxies is closer to Sb-Sc, the inner dynamical masses of the simulations are too high, much like the \citet{Brook2012} simulation.  The \citet{Stinson2013} simulations cover a narrow range in mass, so it is unclear how their feedback prescription would fare for low mass galaxies.

The baryonic feedback models that are being developed to address observational constraints are making headway towards matching observed galaxies.  There is a tendency for the total inner dynamical masses of the simulations to be larger than those of  observed galaxies of similar total stellar mass, with the details depending on the particulars of each simulation.  So far, we have only been able to check the dynamical mass at a particular radius for those simulations that provide the necessary information.  A much stronger test could be made if the radial distribution of baryonic surface density and gravitational force were provided for simulated galaxies \citep{mcgaugh2004}.

\section{Summary}
\label{summary}
We have shown that for a sample of dSph and early- and late-type disk galaxies, the total (baryons+dark matter) dynamical mass inside of 500~pc is nearly independent of the overall mass of the galaxy.  There is a weak scaling, $M(<500\mathrm{pc}) \propto M_{stars}^{0.16}$.  Over $\sim$~9 decades in baryonic mass and $\sim$~3 in length, the central mass increases only 1.5 dex.

The earliest type (S0-Sa) disks are an exception to this weak scaling.  Their central masses increase 1.5 dex rather suddenly for $M_{stars} \gtrsim 10^{10}\;M_{\sun}$.  We speculate that this might be the consequence of a distinct formation mechanism for their dense stellar bulges.

We have also compared the results of recent galaxy simulations to the data to investigate how well the simulations reproduce observed galaxy structure.  The simulations display a slow increase in total inner dynamical mass with increasing total stellar mass that is similar to the trend seen in observed galaxies, though the degree of detailed agreement varies with individual models.  Considerably stronger tests could be made if more information were provided by the models.

\acknowledgements
We would like to thank Matt Walker for sharing his data. We also thank the referee for comments that helped to improve the galaxy sample and sample of comparison simulations. The work of S.S.M. is supported in part by NASA ADAP grant NNX11AF89G.  We acknowledge the usage of the HyperLeda database (http://leda.univ-lyon1.fr).



\end{document}